\documentclass[aps,prb,reprint,amsmath,amssymb,a4paper,floatfix,showpacs,citeautoscript,superscriptaddress]{revtex4-1}
\usepackage{mathptmx}
\usepackage[utf8]{inputenc} 
\usepackage{graphicx}
\usepackage{epstopdf}
\usepackage{xspace}
\usepackage[
,textwidth=17.5cm 
,textheight=23cm 
,verbose
,dvips
]{geometry}

\begin{document}

\title{Direct experimental determination of the spontaneous polarization of GaN}

\author{Jonas Lähnemann}
\email{laehnemann@pdi-berlin.de}
\author{Oliver Brandt}
\author{Uwe Jahn}
\author{Carsten Pfüller}
\author{Claudia Roder}
\author{Pinar Dogan}
\author{Frank Grosse}
\affiliation{Paul-Drude-Institut für Festkörperelektronik, Hausvogteiplatz 5--7, 10117 Berlin, Germany}
\author{Abderrezak Belabbes}
\author{Friedhelm Bechstedt}
\affiliation{Institut für Festkörpertheorie und -optik, Friedrich-Schiller-Universität Jena, Max-Wien-Platz 1, 07743 Jena, Germany}
\author{Achim Trampert}
\author{Lutz Geelhaar}
\affiliation{Paul-Drude-Institut für Festkörperelektronik, Hausvogteiplatz 5--7, 10117 Berlin, Germany}

\begin{abstract} 
We present a universal approach for determining the spontaneous polarization $P_\mathrm{sp}$ of a wurtzite semiconductor from the emission energies of excitons bound to the different types of stacking faults in these crystals. Employing microphotoluminescence and cathodoluminescence spectroscopy, we observe emission lines from the intrinsic and extrinsic stacking faults in strain-free GaN microcrystals. By treating the polarization sheet charges associated with these stacking faults as a plate capacitor, $P_\mathrm{sp}$ can be obtained from the observed transition energies with no additional assumptions. Self-consistent Poisson-Schrödinger calculations, aided by the microscopic electrostatic potential computed using density-functional theory, lead to nearly identical values for $P_\mathrm{sp}$. Our recommended value for $P_\mathrm{sp}$ of GaN is $-0.022\pm0.007$~C/m$^2$.
\end{abstract}


\maketitle

Crystals with a singular polar axis are pyroelectric, i.~e., they exhibit a spontaneous polarization $P_\mathrm{sp}$ in equilibrium. An important class of materials with this property are semiconductors that crystallize in the wurtzite (WZ) structure. Among these, GaN stands out as the material used for solid-state lighting \cite{brandt_natmat_2006} and power electronics\cite{Ambacher_jap_1999}. While in the former case $P_\mathrm{sp}$ affects both the color and the luminous efficacy of the light emitters \cite{Deguchi_jjap2_1999,*Waltereit_nat_2000}, it is exploited for transistor design in the latter example\cite{Ambacher_jap_1999}.

The physical origin of a non-vanishing $P_\mathrm{sp}$ lies in the singular polar axis of the WZ structure, while its magnitude is determined by the deviation from the ideal tetrahedral coordination of the atoms. In consequence, $P_\mathrm{sp}$ critically depends on the ionic bonding contribution and the crystal field, in particular the difference between the internal cell-parameter $u$ and its ideal value $\frac{3}{8}$ \cite{Bechstedt_prb_2000}. However, $P_\mathrm{sp}$ is inaccessible for an infinite bulk crystal. The microscopic treatment of the polarization relies on the transformation between a reference state and the investigated system \cite{Resta_rmp_1994,*Resta_2007}, which means that only polarization differences are accessible, experimentally as well as theoretically. For $P_\mathrm{sp}$ of WZ materials, the natural choice for such a reference is the zinc-blende (ZB) phase \cite{Posternak_prl_1990,*Tagantsev_prl_1992,*Baldereschi_prl_1992, Bernardini_prb_1997, *Bernardini_prb_2001}. This phase has the same next neighbor configuration and a similar bond-length as WZ \cite{Posternak_prl_1990,*Tagantsev_prl_1992,*Baldereschi_prl_1992}; however, with $P_\mathrm{sp}$ vanishing for symmetry reasons. 

A theoretical determination of $P_\mathrm{sp}$ following this microscopic treatment requires methods of high sophistication such as density-functional theory (DFT). Two conceptual frameworks are used to define $P_\mathrm{sp}$ as the difference from the ZB phase, namely, the electrostatic approach \cite{Qteish_prb_1992} and the Berry phase method \cite{Resta_rmp_1994,*Resta_2007}. Both DFT methods solve the reference problem by considering WZ/ZB heterostructures to determine $P_\mathrm{sp}$. This approach has been implemented for various WZ semiconductors: BeO \cite{Posternak_prl_1990,*Tagantsev_prl_1992,*Baldereschi_prl_1992}, SiC \cite{Qteish_prb_1992}, and ZnO \cite{DalCorso_prb_1994}, as well as the group III-nitrides \cite{Bernardini_prb_1997,*Bernardini_prb_2001,Bechstedt_prb_2000}.
However, the accuracy of DFT is restricted as the value of $u$ predicted by this technique depends on the choice of the exchange-correlation functional, with published values for GaN ranging from 0.3755 to 0.3815 \cite{Bechstedt_prb_2000, Bernardini_prb_1997, *Bernardini_prb_2001}. 

Unfortunately, $P_\mathrm{sp}$ is far more difficult to obtain directly from experiment. The analysis of WZ heterostructures results in the difference of \emph{total} polarization (including the piezoelectric polarization $P_\mathrm{pz}$) between the constituent materials \cite{Suzuki_jjap2_1999,*Leroux_prb_1999,*Cingolani_prb_2000,*Park_apl_2000}. There has been only one attempt, based on an indirect thermodynamic approach, to arrive at an independent estimate of $P_\mathrm{sp}$ \cite{Yan_apl_2009}. Lacking a reliable experimental determination of $P_\mathrm{sp}$, the values recommended in the literature \cite{Vurgaftman_jap_2003} are those obtained theoretically despite their acknowledged uncertainty.

In this Rapid Communication, we deduce $P_\mathrm{sp}$ of GaN from the emission energies of excitons bound to basal plane stacking faults (SFs) of the intrinsic I$_1$ and I$_2$ as well as the extrinsic E type \cite{Stampfl_prb_1998}. These SFs are a local deviation from the hexagonal WZ (0001) stacking sequence to the cubic ZB (111) stacking sequence \cite{Blank_pssb_1964} with a structurally well-defined thickness. In other words, our samples are an experimental implementation of the WZ/ZB heterostructures used to theoretically determine $P_\mathrm{sp}$ \cite{Posternak_prl_1990,*Tagantsev_prl_1992,*Baldereschi_prl_1992}. Due to the smaller bandgap of the ZB modification, these SFs form \emph{perfect} ZB-like quantum wells (QWs) in a WZ matrix \cite{Rebane_pssa_1997,*Salviati_pssa_1999,*Sun_jap_2002,*Liu_apl_2005,*Paskova_procspie_2006,Jacopin_jap_2011}, in the sense that fluctuations in composition and thickness do not exist. Moreover, the in-plane lattice constants of the two modifications are close to each other ($\Delta a/a < 2 \times 10^{-3}$) \cite{Brandt_1998}, resulting in a negligibly small contribution of $P_\mathrm{pz}$. In consequence, the internal electrostatic field in the QWs formed by the SFs is \emph{directly} given by $P_\mathrm{sp}$ in the WZ matrix. We use the observed \emph{differences} in transition energies between the SF-types in conjunction with a parameter-free plate capacitor model to determine the strength of the field and thus $P_\mathrm{sp}$. These results are quantitatively confirmed by self-consistent Poisson-Schrödinger calculations for different band alignments. The effective electronic width of the SFs needed for these calculations is obtained from the microscopic electrostatic potential computed by DFT within the superlattice approach. In this work, we realize the theoretical physical concept of WZ/ZB heterostructures \cite{Posternak_prl_1990,*Tagantsev_prl_1992,*Baldereschi_prl_1992,Bechstedt_prb_2000, Bernardini_prb_1997, *Bernardini_prb_2001,Qteish_prb_1992,DalCorso_prb_1994,Resta_rmp_1994,*Resta_2007} experimentally in order to quantify $P_\mathrm{sp}$ for wurtzite semiconductors; for simplicity, we will refer to the value obtained following this definition as $P_\mathrm{sp}$ of GaN.

\begin{figure}
\includegraphics*[width=8cm]{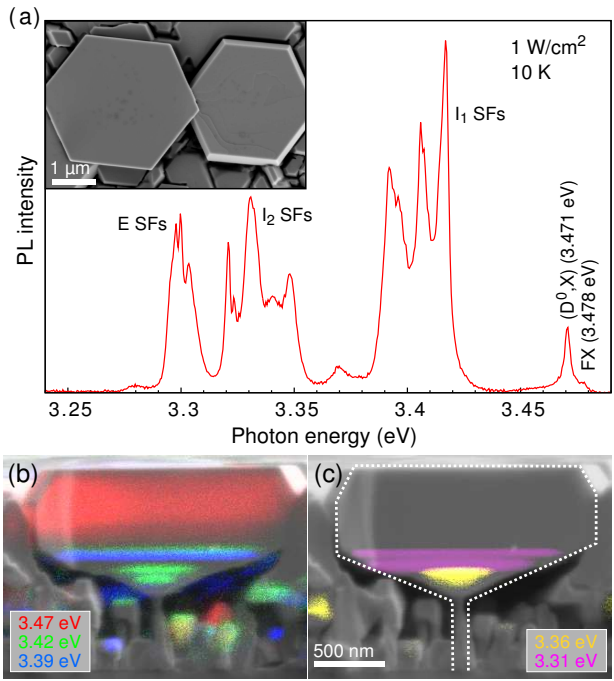}
\caption{\label{fig:1} (Color) (a) $\mu$PL spectrum of an individual microcrystal at low-excitation density. (D$^0$,X) and FX are at the position of strain-free material, while individual SFs and SF bundles emerge as sharp lines right below the energies of the three SF types. The inset shows a top-view scanning electron micrograph of such microcrystals. (b), (c) False-color monochromatic CL maps superimposed on the corresponding cross-sectional scanning electron micrograph of a GaN microcrystal with the detected energies denoted in the respective colors. The CL related to SFs shows a characteristic elongation along the basal plane.}
\end{figure}

The sample used in this study was obtained by pendeoepitaxial overgrowth of self-induced GaN nanowires \cite{Dogan_jcg_2011,*Dogan_cgd_2011} using plasma-assisted molecular-beam epitaxy. The resulting GaN microcrystals are unique with respect to two important properties: they are free of strain and dislocations and contain all three types of SFs. For the spectroscopic analysis of individual microcrystals, we employed a Gatan Mono-CL3 cathodoluminescence (CL) system and a Jobin-Yvon microphotoluminescence ($\mu$PL) setup. The CL system is equipped with a photomultiplier and a charge-coupled device (CCD) detector and mounted to a Zeiss Ultra55 field-emission scanning electron microscope \cite{Jahn_prb_2010}. To achieve a high spatial resolution ($\leq 50$~nm), the acceleration voltage was set to 2 or 3~kV, while the spectral resolution was chosen to be in the range of 1--5~meV. For $\mu$PL, we used the 325~nm line of a Kimmon He-Cd laser for excitation \cite{Pfuller_prb_2010} focused to a spot diameter of about 3~$\mu$m and attenuated to an intensity of 1~W\,cm$^{-2}$. The spectral resolution was set to 1~meV. For all measurements, the samples were cooled to 10~K. 

Figure~\ref{fig:1}(a) presents a low-excitation $\mu$PL spectrum of a single GaN microcrystal detached from the film and dispersed onto a Si substrate. At high energies, emission from bound [(D$^0$,X)] and free (FX) excitons is observed at the positions expected for strain-free GaN. The spectrum is dominated by lower energy lines with a high-energy onset at about 3.42, 3.35 and 3.30~eV, which is associated with emission from I$_1$, I$_2$, and E SFs, respectively. Monochromatic CL imaging on the cross section of such a crystal, as depicted in Figs.~\ref{fig:1}(b) and \ref{fig:1}(c), shows that the near-band edge luminescence originates from the upper part of the microcrystals, while the remaining lines originate from the (lower) region of lateral expansion from a single nanowire to a microcrystal. All the low-energy lines appear as stripes running along the basal plane, confirming their common origin. 

\begin{figure}[t]
\includegraphics*[width=8cm]{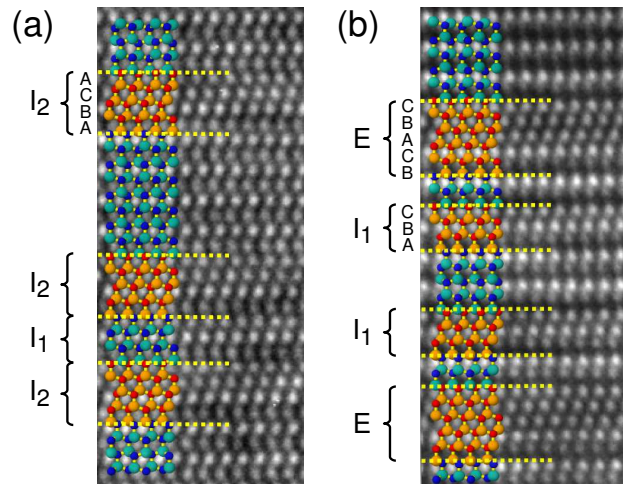} 
\caption{\label{fig:2}(Color) High-resolution transmission electron micrographs revealing SFs of all three types and the bundling of such SFs with only thin WZ interlayers. The micrographs (a) and (b) were taken on two crystals from the same sample used for the spectroscopic analysis. Both micrographs contain I$_1$ SFs. In addition, micrograph (a) contains the I$_2$ and micrograph (b) the E SF. The overlaid ball-and-stick model (note that bright spots correspond to tunneling positions) illustrates the stacking sequence as also indicated explicitly for each SF type.}
\end{figure}

Figure \ref{fig:2} shows high-resolution transmission electron micrographs of this transition region for two GaN microcrystals. All three types of SFs are observed in this region. The images also show that the SFs are often separated by only a few layers of WZ GaN. This ``bundling'' of SFs will modify the transition energy associated with a specific SF arrangement. For larger spatial separations, the electric fields within the aggregate of SFs will be redistributed such as to blueshift the transitions with respect to an isolated SF \cite{Fiorentini_prb_1999}, while a small separation will lead to a coupling of the associated electronic states, resulting in a redshift of the transitions \cite{Paskov_jap_2005}. Hence, the random separation of SFs of different types as observed in Fig.~\ref{fig:2} leads to a (potentially) continuous distribution of the transition energies, explaining the multiple peaks observed in Fig.~\ref{fig:1}(a). The transition energy of isolated SFs is, in contrast, well-defined and is thus expected to dominate statistically.

\begin{figure}[t]
\includegraphics*[width=8cm]{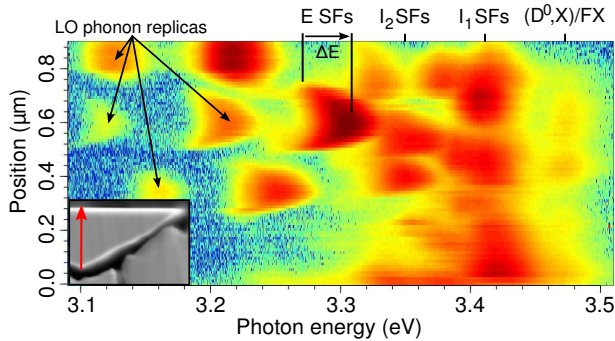} 
\caption{\label{fig:3}(Color) CL spectral image acquired on the cross section of a microcrystal along the arrow indicated in the scanning electron micrograph in the inset. For each position along this line, a spectrum is recorded which is displayed in the spectral image with its intensity color coded on a logarithmic scale. Note that the energy is blue-shifted by $\Delta E$ (26~meV for the marked case) due to a screening of the polarization field when the electron beam approaches the SFs.}
\end{figure}

Prior to compiling such a statistics, it is crucial to ensure that the internal electrostatic fields we are intending to monitor are not screened due to high excitation densities. The CL spectral image presented in Fig.~\ref{fig:3} shows that screening may indeed occur: the crescent-shaped features reflect that the transition energy is lowest with the electron beam far off the SFs but steadily increases with the electron beam closing in. The maximum blueshift is observed with the electron beam situated directly on the SF. The origin of this behavior is the increase in carrier density with increasing proximity of the electron beam due to diffusion and eventually direct excitation. 

\begin{figure}[b]
\includegraphics*[width=8cm]{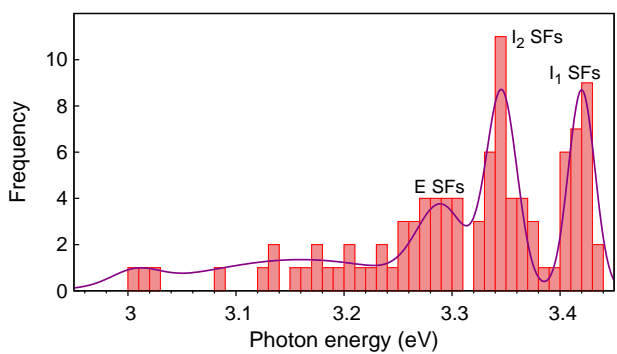}
\caption{\label{fig:4} (Color online) Histogram compiled of peak energies from low-excitation $\mu$PL and of the lowest (unscreened) energies in CL (excluding phonon replicas). The Gaussian fit further illustrates the energies around which the peaks assigned to the different SF types are grouped.}
\end{figure}

To rule out excitation-induced shifts in the determination of the transition energies of the isolated SFs, we thus use exclusively (i) low-excitation $\mu$PL spectra such as the one depicted in Fig.~\ref{fig:1}(a) and (ii) the low-energy onsets obtained for each of the transitions in CL spectral images. Figure \ref{fig:4} shows a histogram of SF-related emission energies compiled from spectra and spectral images recorded on several dozen microcrystals. The energies associated with the isolated SFs emerge in this statistical analysis of the peak energies: we observe three peaks at $3.42 \pm 0.01$, $3.35 \pm 0.01$ and $3.29 \pm 0.015$~eV in good agreement with the high-energy peaks for the three groups of lines in Fig.~\ref{fig:1}(a).\cite{*[{Apart from the bundling of SFs, the error margins may contain additional effects such as localization of electrons by donors in the vicinity of the SFs [}] [{].}] Corfdir_prb_2009} In order of appearance, these peaks are associated with the I$_1$, I$_2$, and E SFs, respectively, as expected from the thickness of the SFs (ZB segments) increasing in this order. While this confirms literature reports on the emission associated with the I$_1$ and I$_2$ SFs \cite{Rebane_pssa_1997,*Salviati_pssa_1999,*Sun_jap_2002,*Liu_apl_2005,*Paskova_procspie_2006}, we also show that luminescence around 3.29~eV can be clearly attributed to excitons bound to E SFs. Emission over the same energy range as displayed in Fig.~\ref{fig:4} was recently reported for faulted GaN nanowires \cite{Jacopin_jap_2011}. The observation of emission below the band gap of bulk ZB GaN was attributed to the presence of internal electric fields, but the limited statistics did not allow for the identification of transitions from isolated SFs and thus prevented an independent estimate of $P_\mathrm{sp}$. The accurate determination of the transition energies from all three types of SFs [cf.\ Fig.~\ref{fig:4}] enables us to go further as detailed in the following.

The crucial point in our analysis is that we rely on the differences in the emission energies rather than on their absolute values. The difference in thickness between each of the SFs is a single (111) bilayer (or molecular monolayer) of ZB GaN, i.~e., $\Delta d = 0.259$~nm. The differences in the emission energies are $\Delta E_{\mathrm{I}_1\to\mathrm{I}_2}=70 \pm 15$~meV and $\Delta E_{\mathrm{I}_2\to\mathrm{E}}=60 \pm 18$~meV; thus the mean energy difference for adding a bilayer to the QW is $\overline{\Delta E} = 65 \pm 23$~meV. Now let us assume that the spontaneous polarization is sufficiently strong so that the single particle energies (relative to the respective band) are governed by the triangular potential and thus remain basically the same for all SFs, while the change in confinement remains negligible. In this case, the polarization sheet charges at the interface between the ZB and WZ modifications essentially represent a plate capacitor, for which the addition of a slab of dielectric of the width $\Delta d$ results in the potential difference $\Delta V = \overline{\Delta E}$. From these elementary considerations, we can directly calculate the polarization sheet charge density
\begin{equation}\label{eq:2}
  \sigma = |P_\mathrm{sp}| = \frac{\Delta V \epsilon \epsilon_0} {\Delta d} = 0.021 \pm 0.007~\mathrm{C/m}^2,
\end{equation}
where $\epsilon=9.5$ is the static dielectric constant for GaN and $\epsilon_0$ is the permittivity of free space. Note that the corresponding electric field within the SF amounts to 2.5~MV/cm.

\begin{figure}[t]
\includegraphics*[width=8cm]{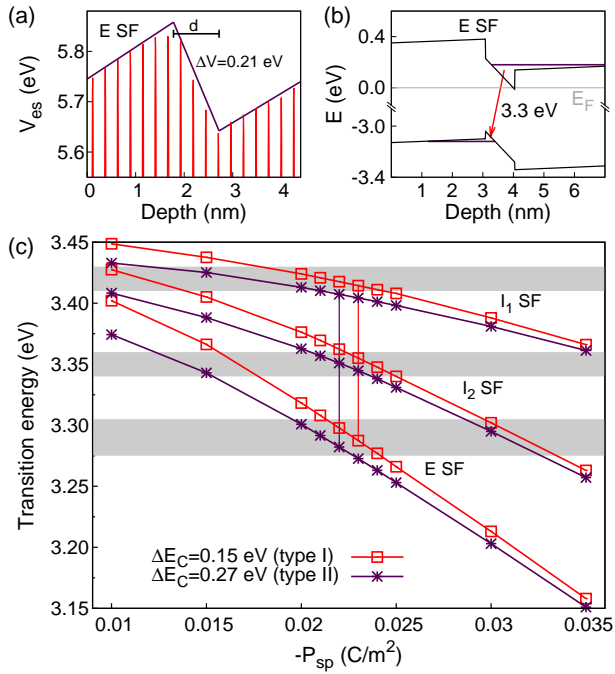}
\caption{\label{fig:5}(Color online) (a) Microscopic electrostatic potential $V_\mathrm{es}$ for the E SF calculated by DFT. From the envelope of the potential, we estimate the thickness (3.7 bilayers) and $\Delta V=0.21$~eV resulting from the polarization field. (b) Band-profile for the E SF from a selfconsistent Poisson-Schrödinger calculation for a type-I band alignment. The horizontal lines denote the single particle energy levels, the arrow the resulting transition. (c) Transition energies for all three SF types from the Poisson-Schrödinger calculations with varied $P_\mathrm{sp}$ and for two different conduction band offsets $\Delta E_C$ corresponding to a type-I and a type-II alignment. The vertical lines denote the values of $P_\mathrm{sp}$ for which the spectroscopically observed \emph{differences} in the transition energy between the SF types are reproduced best; the \emph{absolute} emission energies from experiment (including error bars) are shaded in gray for comparison.}
\end{figure}

The same argument also applies to the microscopic electrostatic potential calculated in the present work and depicted in Fig.~\ref{fig:5}(a) for the E SF (for details, see Ref.~\cite{Belabbes_prb_2011}). Tracing the triangular envelope of the potential in one supercell [cf.\ Fig.~\ref{fig:5}(a)], we estimate an energy difference $\Delta V = 0.21$~eV for a thickness of $d_\mathrm{ESF}=0.96$~nm or 3.7 bilayers. Using Eq.~(\ref{eq:2}) this yields a value of $P_\mathrm{sp}=-0.018$~C/m$^2$. The difference of this value compared to an earlier computation \cite{Bechstedt_prb_2000} is due to
the use of the projector-augmented wave method in the present work, leading to a more accurate description of the electron density near the cores.

To confirm the validity of the assumption inherent in the plate capacitor model used above, we performed self-consistent Poisson-Schrödinger calculations \cite{1DPoisson} for the QWs formed by the I$_1$, I$_2$ and E SFs. We set the effective electronic thickness of the E SF to 3.7 bilayers as motivated from the DFT results displayed in Fig.~\ref{fig:5}(a). The thicknesses of the I$_2$ and I$_1$ SFs were then taken to be 2.7 and 1.7 bilayers, respectively, in accordance with their differences in structural thickness. In order to examine the extent to which the transition energy depends on the respective band offsets, we considered the two extremes suggested in the literature, namely, $\Delta E_C=0.27$~eV (Ref.~\onlinecite{Stampfl_prb_1998}) and $\Delta E_C=0.15$~eV (Ref.~\onlinecite{Belabbes_prb_2011}), resulting in a type-II and a type-I band alignment, respectively. Standard values for the band gaps (containing a correction for the excitonic nature of the transition)\footnote{The exciton binding energy was taken to be equal to the wurtzite bulk value (26~meV) for all SF types. This assumption is justified since both the impact of confinement and the internal electrostatic field (enhancing and lowering the binding energy, respectively) are weak, primarily due to the fact that the SF thickness is well below the excitonic Bohr radius with the exciton wave function largely extending into the wurtzite matrix. Furthermore, these two effects counteract each other, and both become stronger with increasing SF thickness. We thus expect little change of the exciton binding energy for the three types of SFs.} and effective masses were assumed. The residual donor density in our samples was set to $N_d=5\times10^{16}$~cm$^{-3}$.\cite{*[{This value is an upper limit of the O concentration in our samples as measured by secondary-ion mass spectrometry. Independent optical studies of our samples arrive at even lower values [}] [{].}]  Pfuller_nr_2010} Figure \ref{fig:5}(b) shows an example of the band profile, the ground states, and the resulting transition energy for the E SF assuming a type-I band alignment.\cite{*[{Note that the small electric field in the barriers is a consequence of the low donor concentration and the large separation of SFs chosen in the calculation to exclude proximity effects. At these low doping densities, charge redistribution and Fermi level pinning did not noticeably affect the calculations, even for SFs close to the surface in accordance with analogous calculations for (In,Ga)N/GaN QWs [}] [{].}] Mayrock_prb_2000}

\begin{table}[!t]
\caption{Values for the spontaneous polarization $P_\mathrm{sp}$ derived from theory and experiment. EM stands for effective mass and denotes the self-consistent Poisson-Schrödinger calculations.}
\begin{ruledtabular}
\begin{tabular}{lc}
Method&$P_\mathrm{sp}$ (C/m$^2$)\\
\colrule
DFT (Refs. \onlinecite{Belabbes_prb_2011, Bernardini_prb_1997, *Bernardini_prb_2001}) & $-0.014$ to $-0.034$\\
Bond-orbital model (Ref. \onlinecite{Davydov_psolstat_2009}) & $-0.029$\\
Thermodynamics (Ref. \onlinecite{Yan_apl_2009}) & $-0.022$\\
This work (DFT for E SF) & $-0.018$\\
This work (plate capacitor)&$-0.021\pm0.007$\\
This work (EM, type I)&$-0.023\pm0.007$\\
This work (EM, type II)&$-0.022\pm0.007$\\
\end{tabular}
\end{ruledtabular}
\label{tab:1}
\end{table}

Next, we varied $P_\mathrm{sp}$ across the range of values found in the literature (see Table~\ref{tab:1}) for both band alignments. The resulting transition energies for all three SF types are shown in Fig.~\ref{fig:5}(c). Evidently, the limiting values either under- or overestimate the differences in transition energies. The observed $\Delta E=65 \pm 23$~meV is best reproduced for $P_\mathrm{sp}=-0.023 \pm 0.007$~C/m$^2$ in the case of a type I band-alignment and for $P_\mathrm{sp}=-0.022 \pm 0.007$~C/m$^2$ in the case of a type-II band alignment. The fact that both band alignments result in essentially the same value for $P_\mathrm{sp}$ signify that the assumption made in our initial analysis is valid: the influence of confinement is indeed small compared to that of the electric fields, and the SFs essentially behave as plate capacitors.

All values for $P_\mathrm{sp}$ obtained in this work and available in the literature are summarized in Table~\ref{tab:1}. Considering the uncertainty of the former, we recommend $P_\mathrm{sp}=-0.022 \pm 0.007$~C/m$^2$. This value agrees exactly (probably fortuitously) with the only other experimental estimation of $P_\mathrm{sp}$ as reported in Ref.~\onlinecite{Yan_apl_2009} and is also in fair agreement with that directly computed for the E SF in the present work (cf.\ Table~\ref{tab:1}). 

We have demonstrated that the spectroscopic fingerprint of SFs in GaN allows the determination of the strength of the spontaneous polarization in an inherently parameter-free way. This approach may also be used to determine the spontaneous polarization for other important wurtzite materials such as SiC and ZnO.\vspace{4ex}

We would like to thank Steven C. Erwin (Naval Research Laboratory), Holger T. Grahn, and Henning Riechert for a critical reading of the manuscript. This investigation was partially supported by the German BMBF project \textsc{GaNonSi} (Contract No.\ 13N10255).\clearpage

\end{document}